\documentclass{iau} 
\usepackage{graphicx}

\title[Massive Star Clusters in Dwarf Galaxies] 
{Massive Star Clusters in Dwarf Galaxies}

\author[S{\o}ren S.\ Larsen]   
{S{\o}ren S.\ Larsen}

\affiliation{Department of Astrophysics/IMAPP, Radboud University,\\ Postbus 9010,
NL-6500GL Nijmegen, the Netherlands \\ email: {\tt s.larsen@astro.ru.nl} 
}

\pubyear{2015}
\volume{316}  
\jname{Formation, evolution, and survival of massive star clusters}
\editors{C. Charbonnel \& A. Nota, eds.}
\begin{document}

\maketitle

\begin{abstract}
Dwarf galaxies can have very high globular cluster specific frequencies, and the GCs are in general significantly more metal-poor than the bulk of the field stars. In some dwarfs, such as Fornax, WLM, and IKN, the fraction of metal-poor stars that belong to GCs can be as high as 20\%--25\%, an order of magnitude higher than the 1\%--2\% typical of GCs in halos of larger galaxies. Given that chemical abundance anomalies appear to be present also in GCs in dwarf galaxies, this implies severe difficulties for self-enrichment scenarios that require GCs to have lost a large fraction of their initial masses. More generally, the number of metal-poor field stars in these galaxies is today less than what would originally have been present in the form of low-mass clusters if the initial cluster mass function was a power-law extending down to low masses. This may imply that the initial GC mass function in these dwarf galaxies was significantly more top-heavy than typically observed in present-day star forming environments. 
\end{abstract}

\firstsection 
\section{Introduction: globular clusters and halo stars in different environments}

It is well-known that the specific frequency of globular clusters, $S_N$, varies significantly from one galaxy to another\footnote{$S_N \equiv N_\mathrm{GC} \times 10^{0.4\times(M_V+15)}$ for a galaxy of absolute magnitude $M_V$ that hosts $N_\mathrm{GC}$ GCs}. The overall variation of $S_N$ vs. host galaxy $M_V$ is $U$-shaped, with a broad minimum between $M_V\sim-18$ and $M_V\sim-20$ where $S_N$ is typically of order unity, but reaching $S_N > 20$--30 for some cD galaxies and dwarf galaxies (e.g., \cite[Peng et al.\ 2008]{Peng2008}; \cite[Harris et al.\ 2013]{Harris2013}). This trend may be driven partly by dynamical effects that lead to more efficient cluster disruption at intermediate host galaxy masses (\cite[e.g., Mieske et al.\ 2014]{Mieske2014}) and partly by variations in the ``efficiency'' of GC formation, relative to field stars (\cite[Peng et al.\ 2008]{Peng2008}; \cite[Georgiev et al.\ 2010]{Georgiev2010}). Interestingly, the trend of $S_N$ vs.\ host galaxy mass resembles that of total (dark matter) halo mass vs.\ stellar mass inferred from abundance matching (\cite[e.g.\ Behroozi et al.\ 2010]{Behroozi2010}), suggesting that the mechanisms that drive the two may be related.

In addition to this global relation, there are also variations in GC specific frequency \emph{within} galaxies. This is illustrated, for example, by the difference between the integrated colours of early-type galaxies and the mean colours of their GC populations, with the integrated light being redder (\cite[Forte et al.\ 1981]{Forte1981}; \cite[Larsen et al.\ 2001]{Larsen2001}; \cite[Forbes \& Forte 2001]{Forbes2001}). Direct comparisons of the metallicity distributions for halo RGB stars and GCs confirm that the ratio of GCs to halo stars increases towards lower metallicities (\cite[Harris et al.\ 2007]{Harris2007}).
In the Milky Way, GCs account for 1\%--2\% of the mass of the stellar halo (\cite[e.g.\ Larsen et al.\ 2014a]{Larsen2014a}). By decomposing the integrated light of NGC~1399 and M87 into metal-poor and metal-rich components, \cite[Forte et al.\ (2007)]{Forte2007} found that metal-poor GCs account for about 0.9\% and 2.1\% of the metal-poor stars in these galaxies, which is comparable to the corresponding ratio in the Milky Way. In the Large Magellanic Cloud, the integrated luminosity of the halo has been estimated to be about $M_V=-15.2$ (\cite[Kinman et al.\ 1991]{Kinman1991}), while the old GCs have a total absolute magnitude of $M_V=-10.75$ (\cite[Suntzeff et al.\ 1992]{Suntzeff1992}), or about 1.7\% of the total halo luminosity. While these numbers are based on somewhat heterogeneous data and analyses, they do seem to suggest that GCs in general account for a fairly constant fraction of 1\%--2\% of the total stellar halo mass in most galaxies, albeit probably with some real variation. 

It is interesting to compare these numbers with the corresponding fractions of stellar mass in GCs in dwarf galaxies. Within the Local Group, the highest GC specific frequency is found in the Fornax dwarf spheroidal galaxy, which has $M_V\approx-13.2$ and 5 GCs, i.e. $S_N = 26$. The total stellar mass formed over the lifetime of the Fornax dSph is estimated to be $6\times10^7\, M_\odot$ (\cite[Coleman \& de Jong 2008]{Coleman2008}). The total luminosity of the five GCs is $L_V \approx 3.6\times10^5\,L_{V,\odot}$ (\cite[Webbink 1985]{Webbink1985}), which corresponds to a current mass of $\sim7.2\times10^{5} \, M_\odot$, assuming $M/L_V = 2 \, M_\odot/L_{V,\odot}$ for metal-poor GCs (\cite[Strader et al.\ 2009]{Strader2009}). [We note that dynamical measurements suggest a somewhat higher mass-to-light ratio of $M/L_V \sim 3.5 \, M_\odot/L_{V,\odot}$ for the Fornax GCs (\cite[Dubath et al.\ 1992]{Dubath1992}; \cite[Larsen et al.\ 2012b]{Larsen2012b}).]. About half of the initial cluster mass has been lost due to stellar evolution (\cite[e.g.\ Bruzual \& Charlot 2003]{Bruzual2003}), so the initial stellar mass of the clusters would have been $\sim1.4\times10^6 \, M_\odot$ (neglecting dynamical evolution), or about 2\% of the total stellar mass formed in Fornax. However, as in larger galaxies, the metallicity distributions of the GCs and the field stars in Fornax differ significantly: whereas four of the five GCs have $\mathrm{[Fe/H]}<-2$ (\cite[Larsen et al.\ 2012a]{Larsen2012a}), the field star metallicity distribution peaks at $\mathrm{[Fe/H]}\approx-1$ (\cite[Battalia et al.\ 2006]{Battaglia2006}; \cite[Kirby et al.\ 2011]{Kirby2011}). Leaving out the slightly more metal-rich cluster Fornax 4, the remaining four GCs would have had a total initial mass (again neglecting dynamical evolution) of about $1.2\times10^6\, M_\odot$. Only about 5\% of the RGB stars in Fornax have $\mathrm{[Fe/H]} < -2$; scaling the total stellar mass accordingly yields $\sim3\times10^6 \, M_\odot$, so it is clear that the GCs account for a much larger fraction of the metal-poor stars in Fornax compared to, e.g., the LMC or the Milky Way. A more accurate estimate needs to take into account the lifetimes of RGB stars as a function of metallicity and age and therefore depends on the somewhat uncertain star formation history of the Fornax dSph, but \cite[Larsen et al.\ (2012a)]{Larsen2012a} estimated that 20\%--25\% of the metal-poor stars in the Fornax dSph currently belong to the four GCs. 

What about other dwarf galaxies? Within the Local Group, the single GC in the WLM galaxy appears to account for a similar high fraction of the metal-poor stars there (\cite[Larsen et al.\ 2014a]{Larsen2014a}). In their study of GCs in dwarf galaxies beyond the Local Group, \cite[Georgiev et al.\ (2010)]{Georgiev2010} found several other cases of very high specific frequencies, including the \emph{IKN} dwarf galaxy (a member of the M81 group). These authors quote a visual magnitude of $M_V=-11.5$ for IKN, and with 5 GCs the specific frequency is then an astounding $S_N = 125$. The total integrated luminosity of the GCs corresponds to $M_V= -9.3$, so the GCs then account for about 12\% of the \emph{total} luminosity of IKN. From photometry of RGB stars, the field star metallicity distribution appears similar to that in Fornax, with a broad peak around $\mathrm{[Fe/H]}\sim-1$ and only a few stars below $\mathrm{[Fe/H]}\sim-2$ (\cite[Lianou et al.\ 2010]{Lianou2010}; \cite[Tudorica et al. 2015]{Tudorica2015}), so the ratio of metal-poor GC stars vs.\ field stars at low metallicities in IKN is likely to be even higher than in Fornax. It is possible, however, that the luminosity of the IKN galaxy has been significantly underestimated as deep images show it to be very extended (\cite[Okamoto et al.\ 2015]{Okamoto2015}).

Before drawing sweeping conclusions about the clusters, the possibility that these dwarf galaxies may have lost significant fractions of their metal-poor field stars (albeit while holding on to the GCs) must be considered. Neither of them is obviously involved in interactions with any other nearby galaxies, and WLM in particular is located on the outskirts of the Local Group, more than 900 kpc from both the Milky Way and M31. The possible mass loss history of the Fornax dwarf has been studied in considerable detail and from analysis of its luminosity profile, which drops to very low levels well within the tidal radius, \cite[Pe{\~n}arrubia et al.\ (2009)]{Penarrubia2009} concluded that Fornax is unlikely to have lost any field stars. More recently, from modelling of the orbit of Fornax around the Milky Way, \cite[Battaglia et al.\ (2015)]{Battaglia2015} reached a similar conclusion and found that essentially no stellar mass has been lost even for the most eccentric orbits allowed by current observational constraints. IKN is less well studied, but in the stellar density maps of \cite[Okamoto et al.\ (2015)]{Okamoto2015} the galaxy has a rather regular, round shape without hints of tidal tails.

In summary, GCs can account for a much higher fraction of the metal-poor stars in dwarf galaxies (up to 20\%--25\% or more), compared to larger galaxies where the corresponding fraction is typically on the order of 1\%--2\%. It appears unlikely that this difference can be attributed simply to a preferential loss of metal-poor field stars in the dwarfs, so it must reflect differences in the efficiency of cluster formation and/or disruption in the different environments. 
We will now proceed to examine some of the consequences. 

\section{Implications for the dynamical evolution of clusters and cluster systems}

The high GC- to field star ratios have potentially important implications for the amount of mass that GCs could have lost during their lifetimes, either via ``traditional'' dynamical mass loss mechanisms such as two-body relaxation and tidal shocks, or as required in models that seek to explain the presence of chemical abundance anomalies in GCs via self-pollution from AGB or massive stars. 

\subsection{Multiple populations and the mass budget problem}

It is, by now, well established that a large fraction of the stars in globular clusters have ``anomalous'' abundances of many light elements. Compared to stars in the field, some elements are enhanced in these stars (e.g., He, N, Na) while others are depleted (e.g., C, O). Far less well established is the origin of these anomalies, although most current scenarios attribute them to some form of self-enrichment within the clusters with an initial ``pristine'' population of stars (with abundances similar to those seen in field stars) and an ``enriched'' population.  Heavier ($\alpha$- and Fe-peak) elements do not generally vary appreciably between the pristine and enriched populations, and the process responsible for producing the abundance variations is generally thought to be H-burning at high temperatures ($T>20\times10^6$ K; \cite[Langer et al.\ 1993]{Langer1993}). 
Two main classes of polluters have been proposed: massive AGB stars undergoing hot bottom burning (\cite[Ventura et al.\ 2001]{Ventura2001}; \cite[D'Antona \& Ventura 2007]{DAntona2007}) or massive main sequence stars, either fast-rotating (\cite[Prantzos \& Charbonnel 2006]{Prantzos2006}; \cite[Decressin et al.\ 2007]{Decressin2007}) or in binaries (\cite[de Mink et al.\ 2009]{DeMink2009}). 

A common challenge in self-enrichment scenarios is to explain the observed large fractions of enriched stars ($>50$\%). A simple example illustrates this point: in the AGB model, the amount of enriched AGB ejecta produced by the pristine population amounts to about $\beta\approx5\%$ of the initial stellar mass of the cluster for a Kroupa-like IMF (\cite[D'Ercole et al.\ 2008]{DErcole2008}). Thus, even if the enriched gas is turned into new stars with an efficiency of 100\%, it would only be sufficient to form an enriched population with an initial mass of $f_\mathrm{en}=\beta=5\%$ of the initial mass of the pristine population. The ratio of pristine vs.\ enriched low-mass stars would then be 100:5, which is in stark contrast to the observed ratios of 50:50 or even 30:70 (\cite[Bastian \& Lardo 2015]{Bastian2015}). One way out is to postulate that a large fraction of pristine stars were subsequently lost from the clusters; in this basic example $\sim 95\%$ of the pristine stars would have to be lost in order to achieve an equal number of pristine and enriched stars. For every 10 low-mass stars left in the cluster (5 enriched and 5 pristine), there should thus be about 95 ejected stars in the field; in other words, at most $\sim10$\% of the stars that were originally formed in GCs should still be cluster members. There is an evident tension between this calculation and the observation that 20\%--25\% of the metal-poor stars in some dwarf galaxies are members of GCs.

The above calculation is clearly simplistic, and modifications to the AGB scenario are possible that make the numbers look somewhat more favourable. As pointed out in the introduction, about half of the initial mass will have been lost due to stellar evolution after a Hubble time for a standard IMF. If both populations have similar IMFs, they will lose the same fraction of their mass and the final pristine:enriched ratio will be unaffected. However, in the extreme case that the enriched population consists \emph{only} of low-mass stars that are still alive after a Hubble time, the mass budget problem is reduced by a factor of two. Dilution of the polluted ejecta (\cite[e.g., D'Ercole et al.\ 2010]{DErcole2010}) with additional pristine gas may help further. On the other hand, other implicit assumptions made above seem rather unrealistic: 1) star formation efficiencies are likely to be less than 100\%, 2) the calculation assumes that, while a large fraction of the pristine stars are lost, \emph{no} enriched stars escape from the cluster, 3) the comparison with the dwarf galaxies makes no allowance for stars with similar metallicities formed elsewhere, e.g., in the ``field'' or in disrupted (low-mass) clusters. We shall return to this last issue below. One could also imagine invoking a \emph{top-heavy} IMF for the \emph{pristine} population, which would have testable consequences for the production of stellar remnants (\cite[Dabringhausen et al.\ 2009]{Dabringhausen2009}). While the example above is carried out for the AGB scenario, the ``fast rotating massive stars'' scenario has similar difficulties, at least in its basic form (but see \cite[Charbonnel et al.\ 2014]{Charbonnel2014} for possible modifications that may accommodate the GC/field ratios in dwarf galaxies).

Apart from the specific constraints on GC mass loss from dwarf galaxies, there are many other difficulties with self-enrichment scenarios for the origin of abundance anomalies in GCs. We emphasize that there is currently no satisfactory theory, but refer the reader elsewhere for details (\cite[e.g., Bastian \& Lardo 2015]{Bastian2015}; Bastian, these proceedings). 

The above discussion would be void if GCs in dwarf galaxies do not host multiple populations. However, it appears that they do: spectroscopy of a few RGB stars in the Fornax GCs already hinted at the [Na/O] anticorrelation (\cite[Letarte et al.\ 2006]{Letarte2006}), and photometry has shown that they are also likely to exhibit substantial [N/Fe] spreads, comparable to those in Galactic GCs (\cite[Larsen et al.\ 2014b]{Larsen2014b}).


\subsection{Dissolution of low-mass clusters}

The considerations in the previous paragraph pertain specifically to the origin of chemical abundance anomalies in GCs. However, normal dynamical processes such as two-body relaxation and tidal shocks are expected to lead to substantial mass loss from GCs over a Hubble time, in any case. 
Indeed, such dynamical evolution has often been invoked as a mechanism suitable for transforming the power-law mass functions observed in young cluster populations into the mass functions observed in old GC systems, the latter having a characteristic break around $2\times10^5 \, M_\odot$ that shows up as a peak when plotted in logarithmic units (e.g., \cite[Fall \& Zhang 2001]{Fall2001}; \cite[Jord{\'a}n et al.\ 2007]{Jordan2007}; \cite[Kruijssen \& Portegies Zwart 2009]{Kruijssen2009}). 

\cite[Jord{\'a}n et al.\ (2007)]{Jordan2007} show that an excellent fit to the GC mass function is provided by the ``evolved Schechter function'', given as
\begin{equation}
  \frac{\mathrm{d} N}{\mathrm{d} M} \propto (M + \Delta M)^{-2} \times \exp\left(-\frac{M + \Delta M}{M^\star}\right)
  \label{eq:es}
\end{equation}
where $\Delta M$ is the (average) amount of mass lost per cluster, here assumed to be independent of the cluster mass, which is likely to be an oversimplification (e.g. \cite[Kruijssen \& Portegies Zwart 2009]{Kruijssen2009}). $M^\star$ is an exponential cut-off mass to account for the steepening of the MF at high masses that is also seen in young cluster systems (\cite[Larsen 2009]{Larsen2009}; \cite[Portegies Zwart et al.\ 2010]{Portegies2010}). For $\Delta M = 0$, the function is simply a power-law with an exponential truncation, and if mass loss has occurred we see that the MF is flat for $M< \Delta M$, as observed in old GC systems. A fit to the Milky Way GC system yields $M^\star =8\times10^5 M_\odot$ and $\Delta M = 2.5\times10^5\, M_\odot$, and the same parameters also fit the combined GC mass function for dwarf galaxies in the Local Group well. The amount of mass lost from the cluster system can now be found by integrating Eq.~(\ref{eq:es}) over all masses for $\Delta M = 0$ and $\Delta M = 2.5\times10^5 \, M_\odot$. The result depends on the lower limit adopted for the initial mass distribution; for lower limits of $100 \, M_\odot$ and $5000 \, M_\odot$ (and adopting an upper present-day mass limit of $10^6\,M_\odot$), the ratios of the initial to current masses are $M_\mathrm{init}/M_\mathrm{current} = 24$ and 13, respectively.
This calculation takes does not account for stellar evolution, which would again increase the ratios by about a factor of two. However, for the sake of comparison with present-day stellar populations, we can ignore stellar mass loss as this will have affected the clusters and the field populations about equally.
From a more detailed calculation, \cite[Kruijssen \& Portegies Zwart (2009)]{Kruijssen2009} found factors of 64 and 39 for the same lower mass limits, which do include stellar evolution.

The large fractions of dynamical mass loss calculated above may, once again, be compared with the present-day fractions of metal-poor stars in dwarf galaxies that are GC members. As before, far more stars should have been lost from clusters than those that are observed today in the field. Thus, while it is already difficult to accommodate the expected amount of mass loss, either from standard dynamical evolution or as required by simple self-enrichment scenarios separately, the problem becomes even more acute if both were to have occurred. 

\subsection{Infant mortality and the efficiency of cluster formation}

Finally, we comment on the issue of ``infant mortality''. This term is generally used to refer to an early phase of cluster disruption during which a large fraction of the cluster population is dissolved, more or less independently of mass. The notion dates back at least to \cite[Lada \& Lada (1991,2003)]{Lada1991,Lada2003}, who found that the birth rate of young embedded clusters appears to exceed that of older open clusters by at least an order of magnitude. Subsequent studies have found similar results when comparing compact stellar groupings in external galaxies with older clusters (\cite[e.g., Chandar et al.\ 2010]{Chandar2010}). Other studies, however, have questioned the need for an early phase of ``infant mortality'', arguing that a large fraction of the stellar aggregates present at young ages are not true clusters, but rather unbound associations (\cite[Bastian et al.\ 2012]{Bastian2012}). 

At any rate, it seems clear that only a relatively minor fraction of stars in present-day star forming regions form within  clusters that remain bound on long time scales. The formation efficiency of bound clusters appears to correlate with star formation rate density (\cite[Larsen \& Richtler 2000]{Larsen2000}; \cite[Goddard et al.\ 2010]{Goddard2010}; \cite[Silva-Villa \& Larsen 2011]{SilvaVilla2011}) and is typically 1\%--10\% at the current epoch, although it may have been significantly higher in the early Universe when galaxies were more gas rich (\cite[Kruijssen 2012]{Kruijssen2012}). 

Again, the observations of GCs in dwarf galaxies put strong constraints on the efficiency of cluster formation and/or ``infant mortality'' in these environments. Even if no other cluster disruption mechanisms have been at work, a very large fraction of the early star formation must have occurred in the few, massive star clusters that are still surviving to the present day. It is curious that this situation seems to have changed fairly abruptly for metallicities $\mathrm{[Fe/H]}>-2$.

\subsection{Young massive clusters in dwarf galaxies}

\begin{figure}[b]
\begin{center}
 \includegraphics[width=50mm]{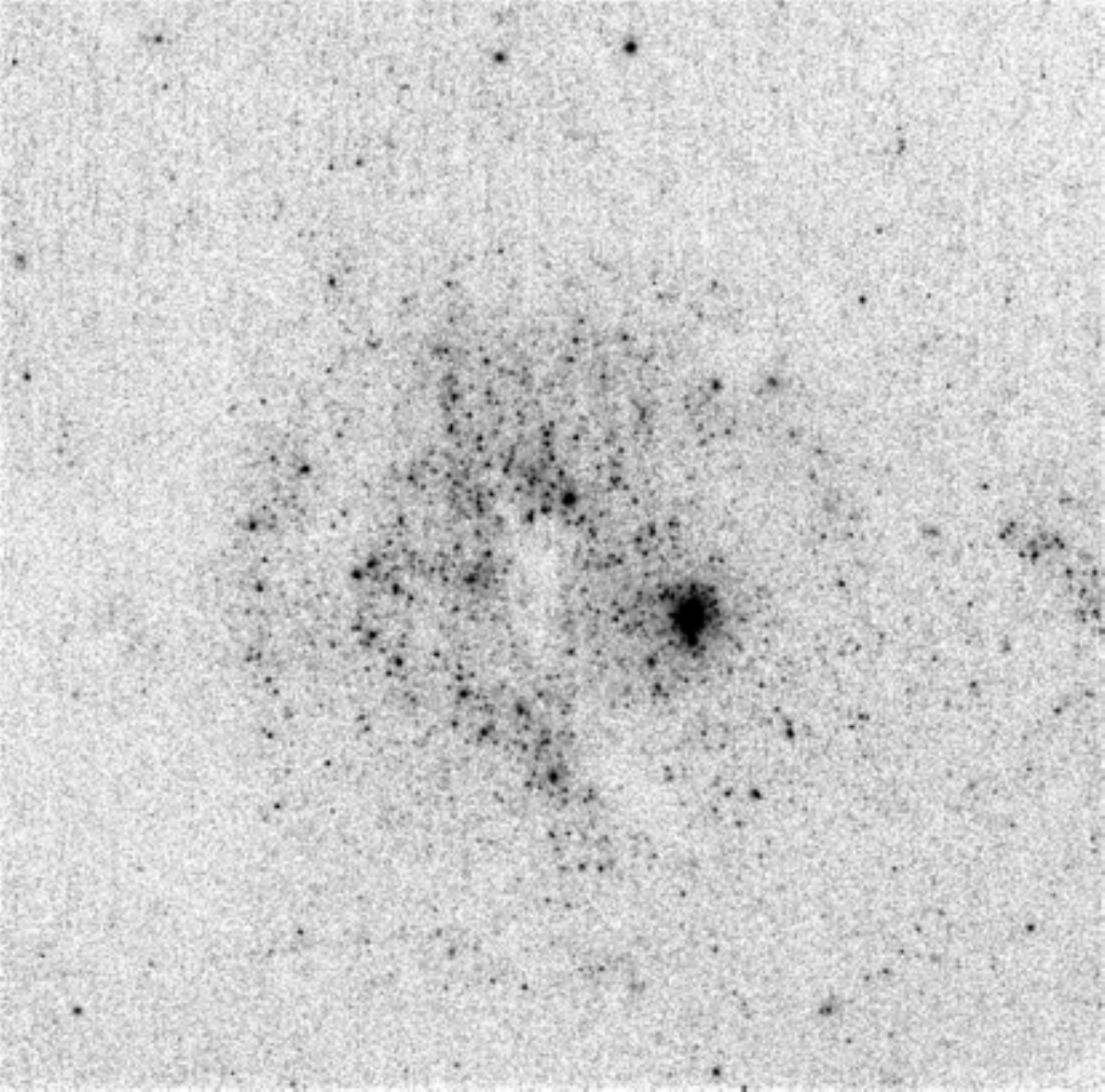}
 \includegraphics[width=50mm]{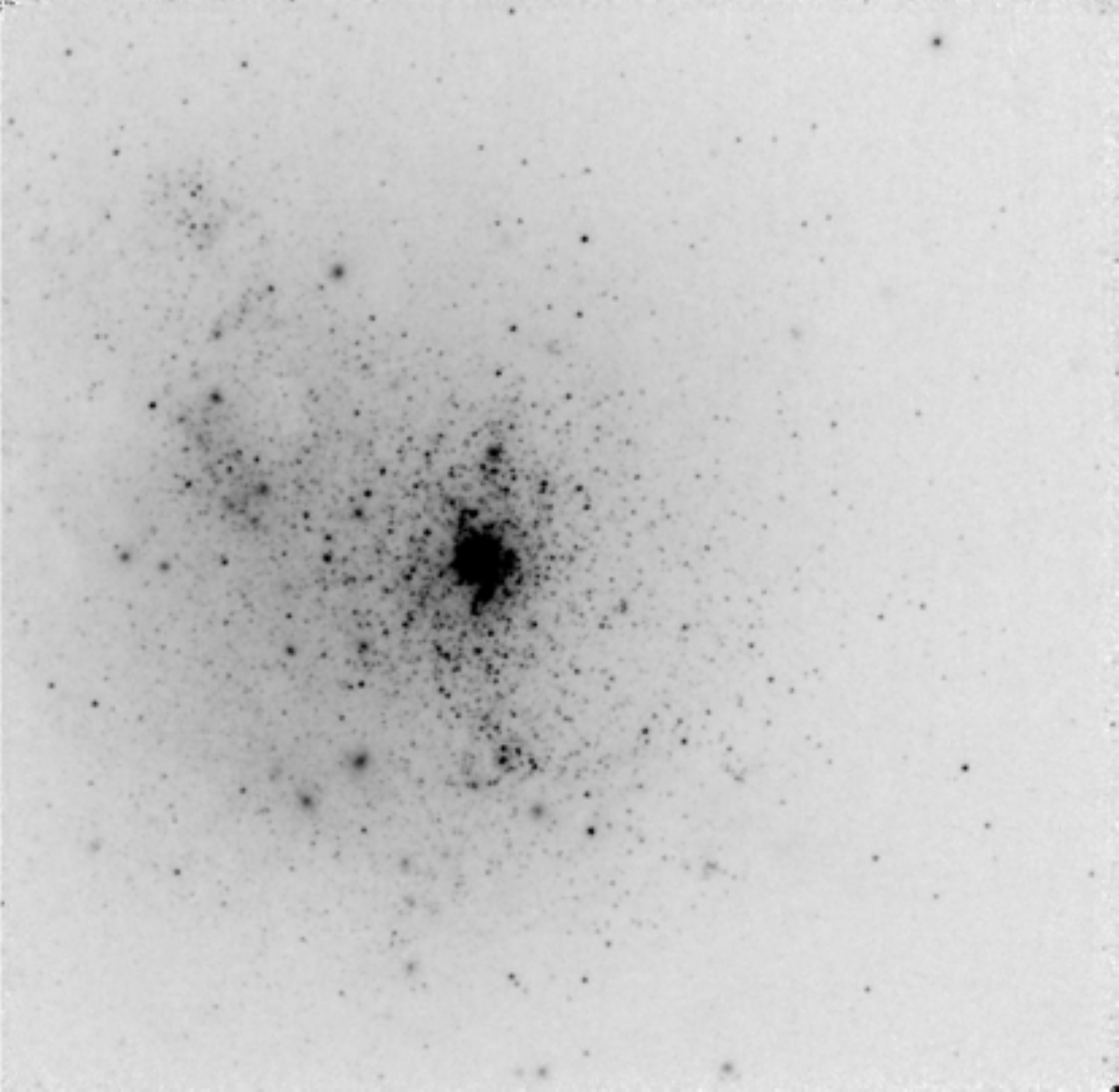}
\end{center}
 \caption{HST/PC F555W images of a star forming complex in the spiral galaxy NGC~6946 (left) and the dwarf galaxy NGC~1705 (right). The dominant object in each panel is a massive star cluster, which is surrounded by numerous fainter clusters and field stars.}
   \label{fig:fig1}
\end{figure}

In the context of dynamical evolution, it is important to note that any direct information about the low-mass end of the initial cluster MF has now been lost, as clusters with initial masses $M < \Delta M$ have disrupted completely. It is, therefore, not strictly necessary to require that a large population of low-mass clusters were present initially, and we can speculate about other possibilities --- although such speculations should, preferably, be guided by observational evidence.

As already noted, young cluster populations in the present-day Universe generally do appear to follow power-law MFs down to low masses (\cite[Elmegreen \& Efremov 1997]{Elmegreen1997}; \cite[Portegies Zwart et al.\ 2010]{Portegies2010}; \cite[Adamo \& Bastian 2015]{Adamo2015}). The best constraints come from galaxies with rich cluster populations, such as mergers and other starbursts, or large spirals with high star formation rates, although the more limited data available for star-forming dwarfs suggest that many of them also have power-law like cluster mass functions (\cite[Billett et al.\ 2002]{Billett2002}; \cite[Seth et al.\ 2004]{Seth2004}; \cite[Cook et al.\ 2012]{Cook2012}). However, there are some notable exceptions to this statement, in particular the dwarfs NGC~1569 and NGC~1705 which are both dominated by one or two very luminous, young clusters (\cite[O'Connell et al.\ 1994]{OConnell1994}). These clusters have estimated masses in the range $8\times10^5\, M_\odot$ to $1.4\times10^6\,M_\odot$ and ages between $5\times10^6$ years and $16\times10^6$ years (\cite[Larsen et al.\ 2011]{Larsen2011}). 

Figure~\ref{fig:fig1} shows HST/PC images of the central regions of NGC~1705 and a stellar complex surrounding a young massive cluster in the spiral galaxy NGC~6946. The two galaxies are at similar distances, 5.1 Mpc for NGC~1705 (\cite[Tosi et al.\ 2001]{Tosi2001}) and 5.9 Mpc for NGC~6946 (\cite[Karachentsev et al.\ 2000]{Karachentzev2000}), so the scales are comparable. The corresponding field-of-view is $0.9\times0.9\,\mathrm{kpc}^2$ and $1\times1\,\mathrm{kpc}^2$.
The similarity of the two regions is striking. Both clusters have masses of about $10^6 \, M_\odot$ and ages of 10--15 Myr and  dominate their immediate surroundings. The cluster in NGC~6946 accounts for about 17\% of the $V$-band light of the surrounding stellar complex (\cite[Larsen et al.\ 2002]{Larsen2002}), while that in NGC~1705 appears to have formed as part of a starburst that produced only $10^6\, M_\odot$ of stars, excluding those in the cluster (\cite[Annibali et al.\ 2009]{Annibali2009}). Indeed, there are no other clusters of comparable brightness and mass in NGC~1705, although the cluster mass function integrated over all of NGC~6946 is consistent with a Schechter-like function.

Perhaps one should imagine the Fornax dwarf and others like it, during the earliest stages of their assembly histories, as collections of localised starbursts like those in Fig.~\ref{fig:fig1}. Within each region, the conditions may have been conducive to forming massive clusters, for example because of high pressures and densities (\cite[Johnson et al.\ 2015]{Johnson2015}). This situation may have applied more generally in the fragments that merged to form galaxy halos, at least those that were large enough to form massive clusters. In this context, it has been suggested that the WLM galaxy, with its single massive GC, may be a surviving analogue of these fragments (\cite[Elmegreen et al.\ 2012]{Elmegreen2012}), although its GC is more metal-poor than typical halo GCs in the Milky Way (\cite[Larsen et al.\ 2014a]{Larsen2014a}).

\section{Summary and Conclusions}

High GC specific frequencies appear to be a common feature in dwarf galaxies. It should, of course, be kept in mind that one GC in a galaxy of absolute magnitude $M_V=-15$ already yields $S_N\equiv1$, so even a single GC will cause a high $S_N$ in faint dwarfs such as Fornax or IKN. The study by \cite[Georgiev et al.\ (2010)]{Georgiev2010} also includes several dwarf galaxies where no GCs have been detected. However, populations of 5 GCs in galaxies like Fornax and IKN seem unlikely to be simply statistical outliers. In any case, we have seen that these high $S_N$ values, combined with the tendency for GCs to be significantly more metal-poor on average than the majority of field stars in their parent galaxies, put tight constraints on the amount of cluster disruption that could have occurred in these galaxies. First, it is difficult to accommodate the large amounts of mass loss required in basic self-enrichment scenarios for the origin of multiple stellar populations in GCs. However, a more general problem is that if the current GC mass function is the result of dynamical evolution from an initial power-law function similar to that observed in young cluster populations, then this should have produced a larger number of field stars than is currently observed, unless the low-mass clusters were preferentially more metal-rich than the surviving GCs. The possibility that the GC mass function was top-heavy already from the outset may thus warrant serious consideration. It appears that a large fraction of the early star formation in these dwarfs may have occurred in the globular clusters that survive to this day.

\section*{Acknowledgements}

I am grateful to the Leids Kerkhoven-Bosscha Fonds (LKBF) and the IAU for travel grants that made it possible for me to attend this symposium.

%
%

\end{document}